\documentstyle[aps,pre,graphicx,amssymb,amsbsy,amsfonts,multicol]{revtex}

\begin{document}

\title{Dynamics of rumor propagation on small-world networks}

\author{Dami\'an H. Zanette}

\address{Consejo Nacional de
Investigaciones Cient\'{\i}ficas y T\'ecnicas\\ Centro At\'omico
Bariloche and Instituto Balseiro\\ 8400 Bariloche, R\'{\i}o Negro,
Argentina}

\date{\today}
\maketitle

\begin{abstract}
We study the dynamics of an epidemic-like model for the spread of
a rumor on a small-world network. It has been shown that this
model exhibits a transition between regimes of localization and
propagation at a finite value of the network randomness. Here, by
numerical means, we perform a quantitative characterization of
the evolution in the two regimes.  The variant of dynamic small
worlds, where the quenched disorder of small-world networks is
replaced by randomly changing connections between individuals, is
also analyzed in detail and compared with a mean-field
approximation.
\end{abstract}


\begin{multicols}{2}

\section{Introduction}
\label{intro}

The  networks  that  underlie  real  social  interactions,  whose
nodes represent single individuals and whose links connect
individuals  that are expected to  interact, vary with  time and
strongly  depend on the kind of  interactions involved.
Generally, however,  social networks exhibit two specific
topological  properties that are closely  related to  the nature
of  social  interactions.   First,  they  are  highly clustered,
which means that two randomly chosen neighbors of a given
individual have  a  relatively  large  probability  of being in
turn mutual neighbors. Second, the distance between any two nodes
in the network, measured as the number of links of the minimal
path connecting the two nodes, is on the average very small as
compared with the total  number of  nodes  or  links. This  is
the  so-called  small-world  effect. Small-world networks (SWNs)
constitute a mathematical model for social networks  that
captures  these  two  properties \cite{swm}. They are partially
disordered networks, interpolating between regular lattices and
fully  random graphs.   In fact,  $N$-node regular lattices with
connections beyond  nearest neighbors have high clustering, but
the average distance between nodes is of order $N$. On the other
hand, the average distance in random networks is of order $\ln N
\ll N$, but the probability that two neighbors of a given node are
mutual  neighbors is of order $N^{-1}$.  For moderate disorder,
SWNs  preserve the two desirable properties of ordered    and
random networks \cite{amar,new,new2,barrat}, and are  therefore a
convenient tool for the mathematical study of social processes
\cite{watts}.

Small-world networks are built  starting from an ordered  lattice
with moderately high  connectivity, which  insures high
clustering. Then, each link is removed with probability $p$ and
reconnected between  two randomly selected nodes. This process
creates a shortcut between  two otherwise distant regions of the
network. The probability $p$ measures the degree of disorder or
{\it randomness} of the resulting graph. For $p=0$ order  is
fully  preserved, while  for $p=1$  a random  graph is obtained.
Note however that the average connectivity is constant.

Topological properties of SWNs,  such as the average  distance
between nodes,  display  a  crossover  from  the behavior
observed in regular lattices to that of random graphs at a
randomness $p\sim N^{-1}$.   In the  limit  of  an  infinitely
large  network,  a critical transition between  both   regimes
occurs   at  $p_c=0$   \cite{mou1,mou2}.    Similar transitions
at the  same  critical  point  are found for some simple dynamical
processes  on  SWNs,  such  as  for Ising-like spin systems
\cite{barrat} and ensembles of  coupled oscillators.  In
contrast, it has  been  recently  shown  that  other  kinds  of
processes display a transition between qualitatively different
dynamical regimes at finite values of the randomness.
Specifically, in an epidemiological  model where  an  initially
susceptible  individual  infected  by  contagion undergoes a
disease cycle  that returns  to the  susceptible state, a
transition at  finite $p$  occurs between  a regime  where the
disease cycles of different individuals are temporally uncorrected
(low $p$) to a  regime  where  the   cycles  synchronize (high
$p$)   \cite{ka}. Moreover,  in  an  epidemic-like model  for
rumor  propagation   a quantitatively  similar transition  has
been  found between a regime where the  rumor remains  localized
(low  $p$) to  a regime  where it spreads over a finite fraction
of the network (high $p$) \cite{z}.

The critical-phenomenon nature of the transition found in the
model of rumor propagation has been convincingly proven by means
of finite-size scaling analysis \cite{z}. This paper, on the
other hand, focuses on a detailed characterization of the
dynamical properties of the same model, with emphasis on the
effects introduced by the small-world topology.  In the next
section we introduce the model and summarize the main results on
the critical transition between  the regimes of localization and
propagation. The core of the paper, Sects. \ref{infected} and
\ref{DSW}, is devoted to establish the connection between the
several parameters of our model and suitable quantities that
characterize its evolution. This is done both in quenched
small-world networks and in the so-called dynamic small worlds,
where quenched disorder in the interaction links is replaced by
stochastic choice of the interaction partners. We emphasize
similarities and differences between both cases. Finally, we
summarize and discuss the main results.

\section{Model of rumor propagation}
\label{model}

Consider a population  formed by $N$  individuals where, at  each
time step, each  individual adopts  one of  three possible
states. In  the first  state,  the  individual  has  not  heard
the rumor yet. In the second state, the individual is aware of
the rumor and is willing  to transmit it. Finally, in the third
state, the individual has heard the rumor but has lost the
interest  in it, and does not transmit  it. By analogy  with SIR
epidemiological  models \cite{Murray}, these three states are
respectively called susceptible, infected, and  refractory. At
the beginning, only one individual is infected and all the
remnant population  is  susceptible.  The  dynamical  rules  act
as   follows \cite{frau}. At each time step, an individual $i$ is
chosen at  random from  the  infected  population.  This
individual contacts one of her neighbors,  say  $j$.   If  $j$
is  in  the  susceptible  state, $i$ transmits the rumor and $j$
becomes infected.  If, on the other hand, $j$ is already infected
or refractory, then $i$ loses her interest  in the rumor and
becomes refractory.

In qualitative terms,  the dynamics can  be summarized as
follows. In the first stage of the  evolution, the number of
infected  individuals increases and,  at a  lower rate,  the
refractory  population grows as well. As a consequence,  the
contacts of infected  individuals between themselves  and  with
refractory  individuals  become  more frequent. Eventually, the
infected population  begins to  decline and vanishes, and the
evolution stops. At the end, the population is divided into  a
group of $N_R$ refractory individuals, who have been infected at
some stage during the evolution, and a group of susceptible
individuals who have never heard the rumor.  It has been shown
\cite{random} that, in the case where contacts can be established
between any two individuals in the population, the fraction
$r=N_R/N$ of refractory individuals at the end  of the  evolution
approaches  a well-defined  limit $r^*$ for asymptotically large
systems, $N\to \infty$. This fraction is given by the nontrivial
solution to the transcendental equation
\begin{equation}    \label{r*}
r^*=1-\exp (-2r^*),
\end{equation}
i.e.  $r^*  \approx  0.796$.  In  other  words,  some  $20  \%$
of the population never becomes aware of the rumor.

We are here interested in the case where contacts between
individuals are  established  along  the  links  imposed  by  a
socially plausible structure,  namely  a  small-world network.
As  advanced  in   the Introduction, SWNs are  built starting
from  an ordered lattice,  with one individual at each node.   We
choose a one-dimensional array  with periodic boundary
conditions, where each node is connected to its $2K$ nearest
neighbors, i.   e.  to  the $K$ nearest  neighbors clockwise and
counterclockwise.  Then each  of the $K$ clockwise connections of
each node  $i$ is  rewired with  probability $p$ to a randomly
chosen node $j$, not belonging to the neighborhood of $i$. A
shortcut is thus created.  We avoid  double and multiple links
between  node, and discard realizations  where the SWN becomes
disconnected.

Previous analysis of this  system, focused on the
characterization of its final state, has revealed  that a
critical transition between  two well-differentiated  regimes
occurs  at  a  finite value $p_c$ of the randomness \cite{z}. For
$p<p_c$, the final number $N_R$ of refractory individuals,
averaged  over  many  realizations  of  the  system,  is
independent of the population size. Therefore, as $N \to
\infty$,  the ratio $r=N_R/N$ tends to vanish. In such limit,
only an  infinitesimal fraction of the population becomes aware
of the rumor, which  remains localized in a small neighborhood of
its origin. On the  other hand, for $p>p_c$  the average value
of  $r$ approaches  a constant  as $N$ grows.   Finite-size
scaling  analysis in  the specific  case of $K=2$ shows that, for
asymptotically large systems,
\begin{equation}    \label{crit}
r\sim |p-p_c|^\gamma,
\end{equation}
with $p_c\approx 0.19$ and $\gamma \approx 2.2$. For larger
values  of $K$, the critical randomness  $p_c$ decreases. The
exponent  $\gamma$, in contrast, seems to be universal.

\begin{figure}[h]
\begin{center}
\resizebox{\columnwidth}{!}{\includegraphics{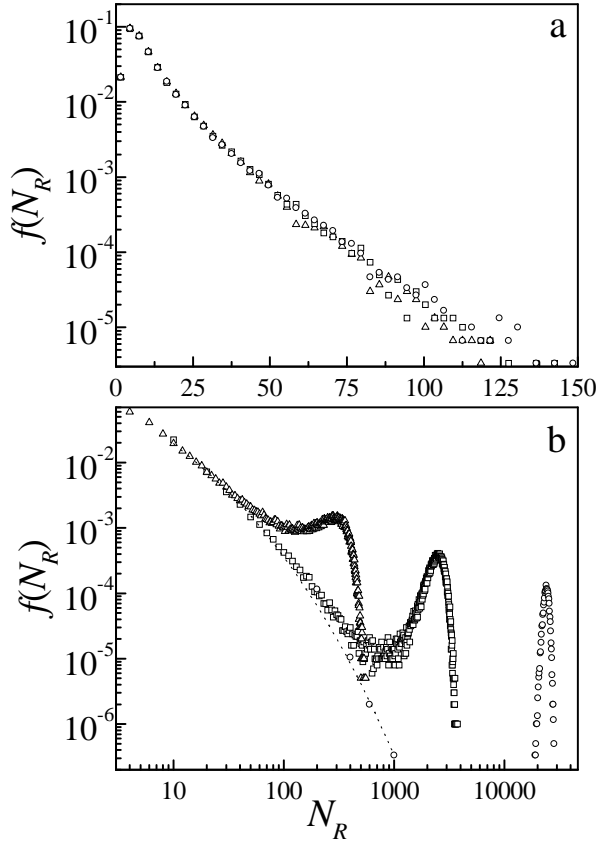}}
\end{center}
\caption{
Normalized frequency distribution $f(N_R)$ of the number of
refractory individuals at  the end  of the  evolution, $N_R$,  for
$K=2$  and two values of the  small-world randomness, (a)
$p=0.05$ and (b)  $p=0.3$.  Different  symbols  correspond   to
$N=10^3$ (triangles),   $N=10^4$ (squares), and  $N=10^5$
(circles). Frequency counts  were  obtained from series of $10^5$
realizations for each parameter set.} \label{f1} \end{figure}

A clue  to the  origin of  the localization-propagation
transition is provided by the  distribution $f(N_R)$ of  values
of $N_R$  over large series of  realizations of  our system  for
fixed  $K$, $p$,  and $N$. Figure \ref{f1} shows those
distributions for $K=2$, two values of $p$---below and above the
transition---and three values of $N$. At  each realization,  the
SWN is constructed anew.   For  $p<p_c$  the  distribution  is
approximately exponential, and does  not depend on  $N$.
Consequently,  the average value of $N_R$ is also independent of
the system size and, as advanced above, the ratio $r=N_R/N$
decreases as $N$ grows. More  specifically, $r\sim N^{-1}$ for
large $N$. In a typical realization for $p<p_c$ the rumor
remains  localized due to  the high interconnectivity of the
network  at  the local level  and  the  scarce density of
shortcuts. Transmission occurs between a small group of
individuals which rapidly loose their interest in the rumor, and
propagation to distant regions is highly improbable.

In contrast, the distribution for $p>p_c$ is bimodal, with two
maxima separated  by  a  local  minimum.   The  small-$N_R$
regime  is still independent of $N$  and attains a  maximum near
$N_R=0$.  On the other hand,  for  large  values  of  $N_R$,  we
find an additional bump-like structure,  which  changes  with
the  system  size. Specifically, the position of its maximum
$N_R^{\max}$ shifts rightward as  $N$ grows, as
$N_R^{\max}\approx  0.25N$. Since,  meanwhile, the  area under
the bump remains  almost constant,  this additional  structure
produces  a contribution  of   order  $N$   to  the   average
value   of   $N_R$. Consequently,  $r$  is  finite  above  the
transition.   While  in  a realization belonging  to the
small-$N_R$ regime  the rumor  remains localized  as  in   the
case  of   $p<p_c$,  a  typical   realization contributing  to
the  bump   includes  propagation  through   several shortcuts,
thus attaining distant regions in the system.

For $p\approx  p_c$, the  distribution (not  shown in  Fig.
\ref{f1}) exhibits power-law  dependence for  moderate values  of
$N_R$, $f(N_R) \sim  N_R^{-\alpha}$  with  $\alpha\approx  1.5$.
The power-law regime terminates  at  a  smooth  cut-off,  whose
position  shifts  to   the right as the system size increases,
approximately as $N^{0.5}$ \cite{z}.

The  localization-propagation  transition  of   our  model  has
been described in  terms of  static features,  namely the  final
refractory population $N_R$, measured when all the interaction
events have ceased. In the next section, we focus on our central
interest here and study the dynamics of the propagation process.

\section{Evolution of the infected population}
\label{infected}

A complete characterization of the propagation process in our
model is given by the evolution of the infected population.
Initially, all  the population is susceptible, except for an
infected individual. Then, at each evolution step, either  the
number of infected  individuals $n_I$ increases to $n_I+1$ at the
expense of the susceptible population,  or $n_I$  decreases  to
$n_I-1$  and  the  refractory  population  grows accordingly.
Therefore,  the   evolution  of  the   number  of   both
susceptible and refractory individuals is implicit in the
evolution  of $n_I$. In order to give $n_I$ as a function of
time, it must be  taken into  account  that  the  number  of
infected individuals varies and, consequently, the real-time
duration of an evolution step changes.  At each evolution step,
in fact, time is to be updated according to
\begin{equation}    \label{time}
t \to t+\frac{t_0}{n_I(t)},
\end{equation}
where the constant $t_0$ fixes time units. We choose $t_0=1$, so
that the unit of time can be  associated with the typical time
needed  by a single infected  individual to  establish a  contact
with  one of  her neighbors.

\vspace{10 pt}

\begin{figure}[h]
\begin{center}
\resizebox{\columnwidth}{!}{\includegraphics{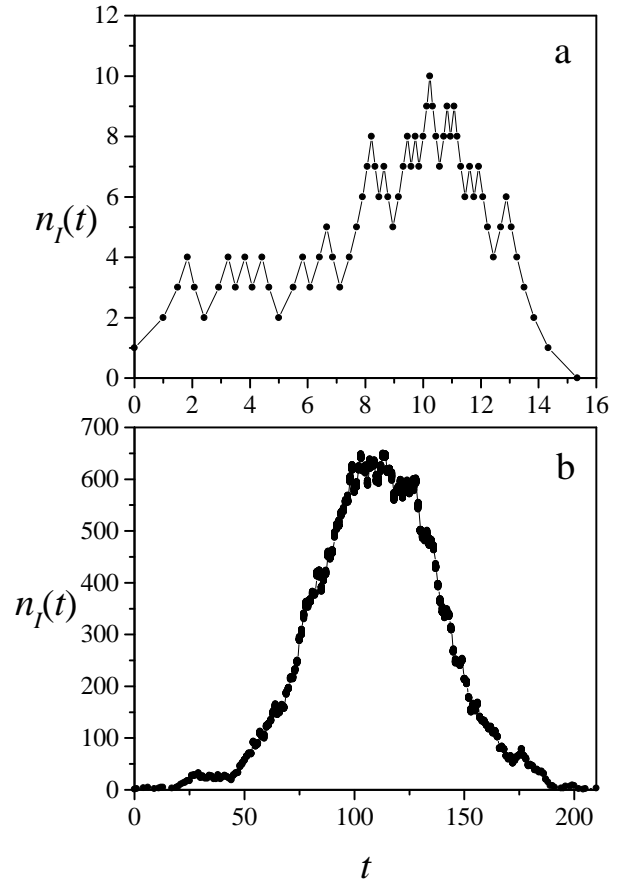}}
\end{center}
\caption{
Evolution of the number $n_I(t)$ of infected individuals as a
function of time in two single realizations on a $10^5$-node
small-world network with $K=2$, for (a) $p=0.05$ and (b) $p=0.3$.
}
\label{f2}
\end{figure}

Figure  \ref{f2}  shows  the  evolution  of  the  number  of
infected individuals as  a function  of time  for two  single
realizations with $N=10^5$ and $K=2$, for two values of the
network randomness, $p=0.05$ and $0.3$. For $p=0.05$ the final
number of refractory individuals is $N_R=32$, while for $p=0.3$
we have $N_R= 22,258$. This latter realization belongs  to the
large-$N_R$ bump  structure in $f(N_R)$ for  the corresponding
value of  $p$. Realizations for the same randomness but in the
small-$N_R$ region are qualitatively similar to that shown in
Fig.  \ref{f2}a.

\vspace{10 pt}

\begin{figure}[h]
\begin{center}
\resizebox{\columnwidth}{!}{\includegraphics{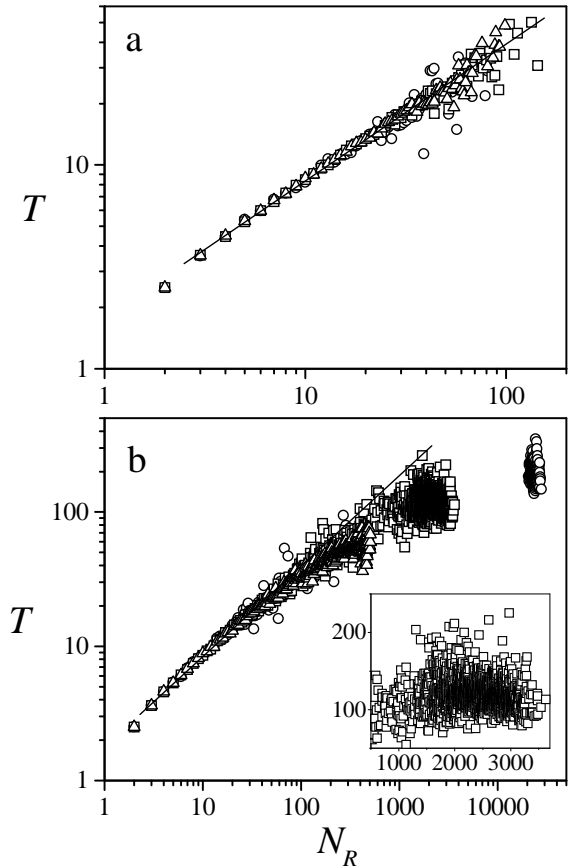}}
\end{center}
\caption{Total  time  $T$  up  to  the  extinction  of  the   infected
population,  as  a  function  of   the  final  number  of
refractory individuals $N_R$, for networks with $K=2$ and
$N=10^3$  (triangles), $10^4$ (squares), and $10^5$ (circles).
The randomness is (a) $p=0.05$ and  (b)  $p=0.3$.  Both  $T$ and
$N_R$  were  measured  in   $10^4$ realizations for each value of
$p$. In the cases where several  values of $T$ were obtained for
the same value of $N_R$, they were  averaged. The straight lines
have slope $2/3$. The insert in (b) shows a close-up of the
``cloud'' at large values of $N_R$ for $N=10^4$, in linear
scales.}
\label{f3}
\end{figure}

The graph  of $n_I$  as a  function of  time for  small $p$---or,
for $p>p_c$, in  the region  of small  $N_R$---is reminiscent of
a random walk. As a  matter of fact,  the evolution of  the
number of  infected individuals can be thought of as a
one-dimensional random walk in  the $n_I$-space, starting  at
$n_I=1$  and  with  an  absorbing boundary condition  at  the
origin,  $n_I=0$,  where the evolution terminates. Equivalently,
we  may  think  of  a first-passage-time problem, with respect to
the origin, for a random walker starting at $n_I=1$  \cite{rw}.
This analogy, however, is difficult to exploit, since in our
case  the random walk would be biased by a complex time-dependent
asymmetry.  In fact, the probability for $n_I$  to grow or
decrease depends  not only on $n_I$  itself,  but  also  on  the
number of both susceptible and refractory individuals.  The
effect of this bias would be particularly strong for the
large-$N_R$ realizations with $p>p_c$.  In this case, indeed, the
evolution of $n_I(t)$ does not resemble a random  trajectory but
mimics deterministic dynamics affected  by a moderate level of
noise (see Fig. \ref{f2}b).

The first-passage-time analogy  suggests anyway that a compact
quantitative characterization of the  propagation process is
given  by the  total  time  $T$  elapsed  up  to  the extinction
of the infected population, and the maximum number of infected
individuals during  the evolution, $N_I$. In the associated
random walk, these two  quantities correspond to  the
first-passage  time and  the maximum  span from the origin,
respectively. In order  to compare with our  previous results,
we  measure  $T$  and  $N_I$  as  a  function  of the final
refractory population $N_R$. Note that $N_R$ is directly related
to the  duration of the propagation process measured in evolution
steps. In fact, since the final number  of infected individuals
is zero, each  step where a susceptible individual becomes
infected must be compensated by a  step where an infected
individual  becomes refractory. Since an  extra step of this
latter kind is  needed for the first infected  individual, the
total number  of steps  necessary for  the extinction  of the
infected population is exactly $2N_R-1$. On the other hand, due
to the changing duration  of  steps  in  real  time,  Eq.
(\ref{time}), the connection between $T$ and $N_R$ is more
complex.

In Fig. \ref{f3}, we present measurements  of the total time $T$
as  a function of $N_R$  over series of  $10^4$ realizations,
for  SWNs with $K=2$ and three different sizes, and for two
values of the  randomness $p$.   For  small  $p$  there  is  a
quite  well  defined   power-law dependence, $T\sim N_R^\tau$,
spanning almost two orders of  magnitude in $N_R$.   Linear
fitting  of these  data yields  an exponent  $\tau$ close to
$2/3$.  This result differs  from the value  predicted by the
random-walk analogy for an  unbiased random walk in  the
$n_I$-space, which gives  $\tau=1/2$.   On the  other hand,  it
can  satisfactorily reproduced by  a random  walk with  constant
bias,  with probabilities $P=0.6$ of  moving towards  $+\infty$
and  $1-P=0.4$ of  moving in the opposite  direction.  We
recall,  however,  that the analogy would be strict for a
time-dependent bias only. For $p>p_c$ the above power-law
dependence is still observed  in the small-$N_R$ regime,  but
apparent deviations  appears  as  $N_R$  grows.   In  particular,
the detached ``cloud'' of dots observed for $N=10^4$ and $10^5$
at large values  of $N_R$---which  correspond  to  realizations
in  the  bump  structure observed  in  $f(N_R)$  (see  Fig.
\ref{f1})---does not satisfy the power-law relation.   The total
evolution  times associated with  such realizations   are
considerably   below   those   predicted   by  an extrapolation
from the small-$N_R$ region, and the difference  becomes larger
as the size $N$ grows.  Note moreover, from the insert in  Fig.
\ref{f3}b,  that  inside  the   ``cloud''  there  is no obvious
correlation between $T$  and $N_R$, in  contrast with the
small-$N_R$ regime. These features  make it evident  that a
qualitative  change in the dynamical behavior occurs between the
regimes of small and  large $N_R$, as illustrated in Fig.
\ref{f2}.

\vspace{10 pt}

\begin{figure}[h]
\begin{center}
\resizebox{\columnwidth}{!}{\includegraphics{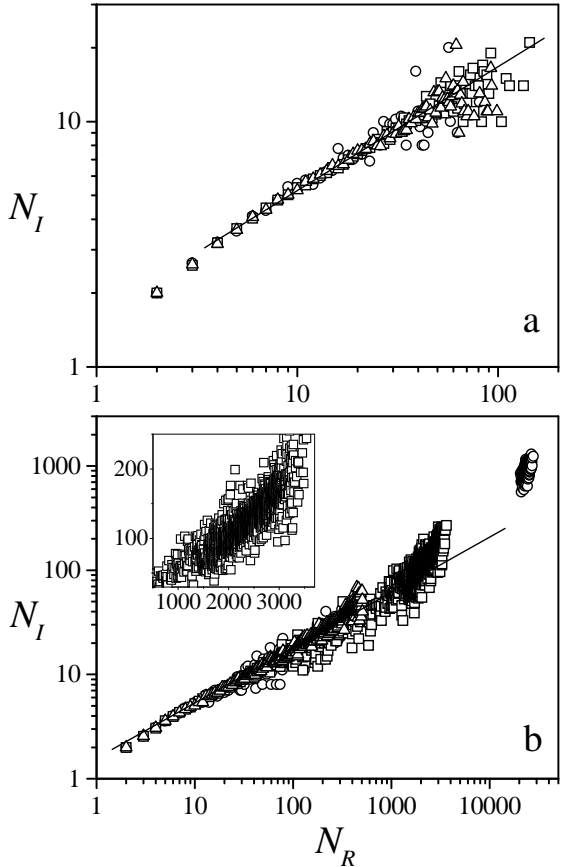}}
\end{center}
\caption{Maximum number of  infected individuals during the evolution,
$N_I$, as  a function  of the  final number of refractory
individuals, $N_R$,  for  networks  with  $K=2$  and  $N=10^3$
(triangles), $10^4$ (squares), and $10^5$  (circles). The
randomness  is (a) $p=0.05$  and (b) $p=0.3$. Data were  obtained
from the same  numerical realizations as those of Fig. \ref{f3}.
The straight lines have slope $1/2$. The insert in (b) shows a
close-up of the ``cloud'' at large values of $N_R$ for $N=10^4$,
in linear scale.}
\label{f4}
\end{figure}

Essentially the same features are found for the dependence on
$N_R$  of the maximum number of infected individuals during the
whole evolution, $N_I$, shown in  Fig.  \ref{f4}.   Now, however,
the  exponent in  the power-law  relation  $N_I\sim  N_R^\nu$,
observed  to  hold  in   the small-$N_R$ regime, is  close to
$1/2$,  which does coincide  with the result for an  ordinary
unbiased random  walk.  For  large $N_R$, the ``clouds'' of dots
quoted above deviate in this case to higher  values of $N_I$.  A
detail of the ``cloud'' for $N=10^5$, shown in the insert of
Fig.  \ref{f4}b, reveals  a remnant  correlation between  $N_I$
and $N_R$.

Note that the power-law dependence of $T$ and $N_I$ on $N_R$,
implies the relation $T\sim N_I^\mu$,  with $\mu=\tau /\nu
\approx  1/3$. This relation is expected to hold for small
randomness or, more  generally, for small $N_R$.

Through the  study of  the evolution  of the  infected
population,  we have so far  examined the dynamical  properties
of our  model for just two values  of the  small-world
randomness  $p$, below  and above  the localization-propagation
transition at $p_c$. It is now worthwhile  to discuss how the
results change as  $p$ is varied. For $0<p<p_c$, as  a matter of
fact, the power-law dependence of $T$ and $N_I$ on $N_R$  is not
modified. The exponents $\tau$  and $\nu$ are the same  within our
numerical  precision.   On  the   contrary,  above   the
transition, substantial changes  affect the  frequency
distribution  of the  final refractory population $N_R$ and its
relation to $T$ and $N_I$.

\begin{figure}[h]
\begin{center}
\resizebox{\columnwidth}{!}{\includegraphics{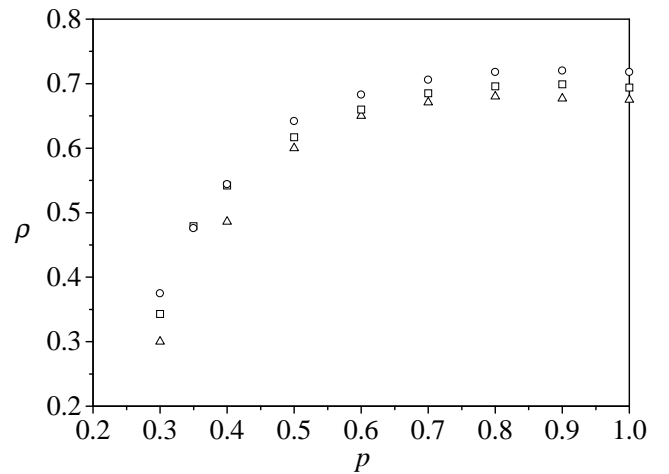}}
\end{center}
\caption{Fraction $\rho$ of realizations in the bump
at large $N_R$ as a function of the small-world randomness $p$,
for networks with $K=2$ and $N=10^3$ (triangles), $10^4$
(squares), and $10^5$ (circles). Data obtained from series of
$10^3$ realizations for each values of $p$.}
\label{f5}
\end{figure}

In  the  first  place,  the  relative  number  of  realizations
in the small-$N_R$ regime and in the bump at large $N_R$ varies
considerably with $p$.  Figure \ref{f5}  shows the fraction
$\rho$ of  realizations in the large-$N_R$ bump as a  function of
$p$ for three values  of $N$ and $K=2$.  This fraction grows from
$\rho\approx0.35$ for $p=0.3$  to $\rho\approx  0.7$  for  the
maximum  randomness  $p=1$. As expected, realizations where the
rumor attains a significant fraction of  the population become
more frequent as the network randomness grows.  Note moreover
that the dependence with  the system size $N$ is  quite weak, but
there is no clear indication of saturation for large $N$.

\vspace{10 pt}

\begin{figure}[h]
\begin{center}
\resizebox{\columnwidth}{!}{\includegraphics{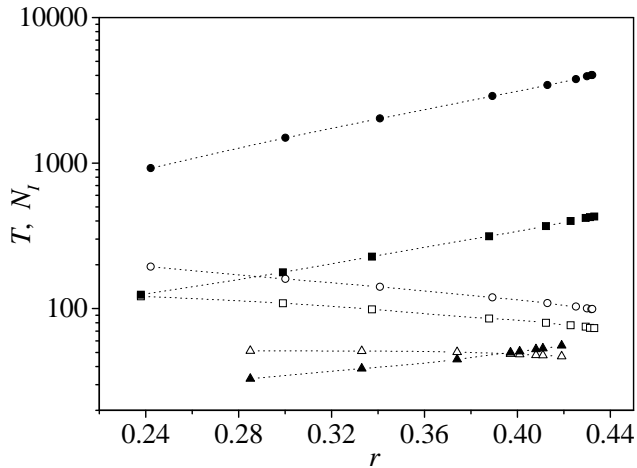}}
\end{center}
\caption{
Average values of the total evolution time $T$ (empty symbols)
and the maximum infected population $N_I$  (full symbols) in the
large-$N_R$ regime as  functions of the average fraction of final
refractory individuals, $r=N_R/N$, for networks with $K=2$ and
$N=10^3$ (triangles), $10^4$ (squares), and $10^5$ (circles).
Dashed lines have been drawn as a guide to the eye. In each data
set, the leftmost and rightmost dots correspond to $p=0.3$ and
$p=1$, respectively. From left to right, the randomness changes
by steps of size $\delta p=0.1$. For $N=10^4$ and $10^5$, the
value $p=0.35$ is also included. Each dot stands for an average
over $10^3$ realizations.}
\label{f6}
\end{figure}

As for the dependence of the total evolution time $T$ and the
maximum number  of  infected  individuals   $N_I$  on  the
final   refractory population  $N_R$,  the  small-$N_R$  regime
exhibits  no significant modifications as $p$ changes.  The
power-law dependence with the  same exponents  is  maintained,
as  expected   from  the  fact  that   the realizations in this
regime correspond to  propagation of the  rumor over a limited
neighborhood of  its origin. In contrast, the  bump in $f(N_R)$
varies in  position and the  corresponding values of  $T$ and
$N_I$ change. In Fig.  \ref{f6}, we show the average values of
$T$ and $N_I$  as  a  function  of  the  fraction  $r=N_R/N$
corresponding to realizations in the bump for several values of
$p$.  Roughly speaking, each  dot  represents  the  centers  of
the ``clouds'' referred to in connection  with  Figs.   \ref{f3}
and  \ref{f4},  now  for   varying randomness.   For  fixed $N$,
the  final refractory population $N_R$ grows  with  $p$. For the
largest  systems,  in  fact,  $N_R$   is practically doubled as
$p$ varies from  $0.3$ to $1$.  We thus  verify again that
propagation is more efficient for larger randomness.  This effect
is enhanced  by the fact  that, at the  same time, the maximum
infected population $N_I$ increases and the total time $T$
decreases. For large systems,  $T$ is reduced  by a factor of $2$
whereas  $N_I$ grows  by  a  factor  of  $5$, approximately. In
summary, the process becomes simultaneously more effective  and
more rapid. Note,  finally, the strong saturation in  the values
of $r$,  $T$, and  $N_I$ as the randomness approaches its maximum
$p=1$.

\vspace{10 pt}

Let us  end this  section by  addressing the  effects of changing
the average number of neighbors per node, given by the parameter
$K$.  As for the localization-propagation transition, a growth in
the number of neighbors implies  that the  critical randomness
$p_c$ decreases and that,  for  fixed  $p$,  the final fraction
of refractory individuals increases \cite{z}. These two  results
agree  with the  expected fact that propagation is more
efficient  for larger $K$. The same  trend is observed  in  the
parameters  that  characterize the evolution of the infected
population: while the  total time $T$ decreases, the maximum
number of infected individuals $N_I$ grows. Our results for
$K=2$,  in any case,  are not  qualitatively changed when other
values of $K$ are considered.

\section{Propagation on dynamic small worlds}
\label{DSW}

Dynamic small worlds (DSWs) have been introduced as a variant to
SWNs in the frame of a model for activity propagation in a system
of mobile automata \cite{manrubia}. Instead  of considering a
frozen disordered interaction  network,   DSWs  admit
interactions  between   any  two individuals occupying the nodes
of a regular lattice.  In the case  of a one-dimensional  array
with  periodic boundary  conditions, at  each interaction event,
the  partner  of  an  individual  is  chosen with probability
$1-p$ among its $2K$ nearest neighbors, $K$ clockwise and $K$
counterclockwise.  With  the  complementary probability $p$, the
partner is chosen at random from  the whole lattice. In this
way,  all individuals  have  the  chance  to interact  with
arbitrarily distant partners, but the probability of distant
interactions is controlled by the ``randomness'' $p$.

The change from SWNs to DSWs, which conveys the replacement of
frozen disorder  by  a  stochastic  process,  is qualitatively
similar to the introduction of the so-called  annealed
approximation in the  study of disordered Boolean (Kauffman)
networks \cite{derrida}. A  considerable advantage  of  DSWs
over  SWNs  regards the numerical implementation, which  does
not  require  the  generation  of  a new lattice at each
realization---a  highly time-consuming  step in  our specific
system. However, the main virtue of DSWs---shared with the
annealed model  for Kauffman  networks---is  that,  in
principle,  they admit a simpler analytical  treatment.   In
particular,  the limit  $p=1$  should be exactly   described, in
asymptotically large   systems,   by   a mean-field-like approach.

We show in the following that the behavior of the rumor
propagation model on a  DSW bears remarkable  similarity with
the  case of a  SWN, though some significant quantitative
differences are detected. Let  us first of all point out that, as
for the dependence of our model on the system size $N$ and on the
average number of neighbors per individual, $2K$, the  features
described  in the  previous section  for SWNs  are qualitatively
reproduced in DSWs.  Consequently, we do not  repeat the analysis
for varying $N$ and $K$, and focus here on the specific  case
$N=10^4$, $K=2$. The behavior for other  values of $N$ and $K$
can  be inferred from this case and the results for SWNs.

\begin{figure}[h]
\begin{center}
\resizebox{\columnwidth}{!}{\includegraphics{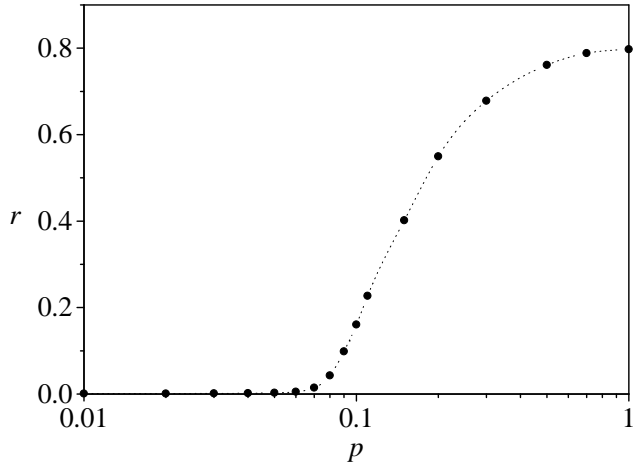}}
\end{center}
\caption{
Final fraction of refractory individuals, $r=N_R/N$, as a
function of the ``randomness'' $p$ on a dynamic small world with
total population $N=10^4$ and $K=2$. Each dot stands for an
average over $10^4$ realizations.  The dashed line is a spline
approximation, drawn as a guide to the eye. }
\label{f7}
\end{figure}

It  has  already  been  advanced  \cite{z}  that  our  model  on
a DSW undergoes the same  kind of localization-propagation
transition found on SWNs.   This is  illustrated in  Fig.
\ref{f7},  where we  show the final fraction of  refractory
individuals $r=N_R/N$  as a function  of $p$, for  $N=10^4$ and
$K=2$.   The critical  point has  considerably decreased, to
$p_c\approx 0.06$. Meanwhile, as expected, the  fraction $r$
approaches the  solution to Eq.  (\ref{r*}), $r^* \approx  0.796$
as $p\to  1$.  Note  in  fact  that  the  original  version of the
model, discussed in Sect.  \ref{model}, is a kind of mean-field
approximation of the small-world case, which becomes exact for
$p=1$.

\vspace{10 pt}

\begin{figure}[h]
\begin{center}
\resizebox{\columnwidth}{!}{\includegraphics{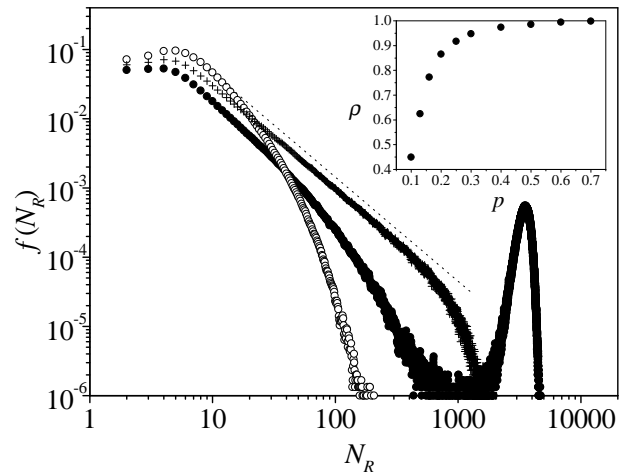}}
\end{center}
\caption{
Normalized frequency distribution $f(N_R)$ of the final number of
refractory individuals $N_R$ on a dynamic small world with a total
population $N=10^4$ and $K=2$, for $p=0.02$ (empty dots), $p=0.06$
(crosses), and $p=0.1$ (full dots). The frequency distribution was
obtained from series of $10^6$ realizations for each
``randomness.'' The dashed straight line has slope $-1.5$. The
insert shows the fraction $\rho$ of realizations belonging to the
large-$N_R$ bump as a function of $p$, averaged over $10^3$
realizations. }
\label{f8}
\end{figure}

As for SWNs, the nature of the localization-propagation
transition  in DSWs is revealed by the frequency distribution of
the final number  of refractory  individuals,   $f(N_R)$.
Figure   \ref{f8}  shows   this distribution for three values of
$p$. For $p=0.02$ we find  a rapidly decaying function which, as
in the subcritical regime on SWNs, results to be roughly
exponential. In this case, in fact, the contribution  of distant
interactions is negligible, so that no significant differences
are  to  be  expected  between  DSWs  and  SWNs.   For $p=0.06$,
which approximately corresponds  to the  critical point  $p_c$
in  DSWs, the distribution is a power law over a substantial
interval, $f(N_R)  \sim N_R^{-\alpha}$.  Remarkably, the exponent
of this power law  coincides---up to the numerical
precision---with that obtained at the  critical randomness  in
SWNs, $\alpha  \approx  1.5$ (cf. Sect. \ref{model}). Finally,
above the critical point  ($p=0.1$), we find that the  by-now
familiar bump  structure at  large $N_R$  has developed.   The
strong similarity with the  scenario on SWNs  convincingly
suggests that  the origin of the localization-propagation
transition is the same for both systems.

The insert of Fig. \ref{f8} shows the fraction $\rho$ of
realizations that contribute to  the bump, as  a function of
$p$.  Comparing  with Fig. \ref{f5},  which shows  the same
results for  SWNs, we note that---apart from the  obvious
consequences of  the shift of  $p_c$ to the left---the fraction
$\rho$  attains  considerably  larger values. In particular, we
find $\rho \approx 1$ for  $0.6\lesssim p$. For such values of
$p$, therefore,  the  rumor  propagates  to  distant regions and
attains  a  finite  portion  of  the  system  in virtually  {\em
all} realizations. An important contribution  to this difference
with  SWNs is given by the following fact. In our SWNs, links
between individuals are bidirectional.  This implies that if  an
infected  individual $i$ transmits  the  rumor  to a susceptible
neighbor  $j$, there is a relatively  high probability  that, in
the future, the now infected individual $j$ will (unsuccessfully)
attempt to transmit the rumor  to $i$ and will become refractory.
These unsuccessful trials for backward propagation are by far
more  improbable in DSWs, especially for large $p$, and the rumor
spreading is consequently enhanced.

\vspace{10 pt}

For $p<p_c$  the total  evolution time  $T$ and  the maximum
number of infected individuals  $N_I$ satisfy  the same power-law
dependence of $N_R$ as in  SWNs. In fact,  as far as long-range
interactions remain infrequent, the evolution on SWNs and DSWs is
essentially  equivalent. The  same  argument  can be  extended
for  $p>p_c$  in   small-$N_R$ realizations.  In this  case,
however,  the  exponent  $\tau$ in the relation $T\sim N_R^\tau$
results to be smaller than for SWNs; now, we find $\tau\approx
0.57$. The moderate contribution of distant contacts is  here
enough  to  produce  a  considerable decrease  of the total
evolution time.  Since contacts  between any  two individuals
are now possible, propagation on DSWs is faster. \vspace{10 pt}

\begin{figure}[h]
\begin{center}
\resizebox{\columnwidth}{!}{\includegraphics{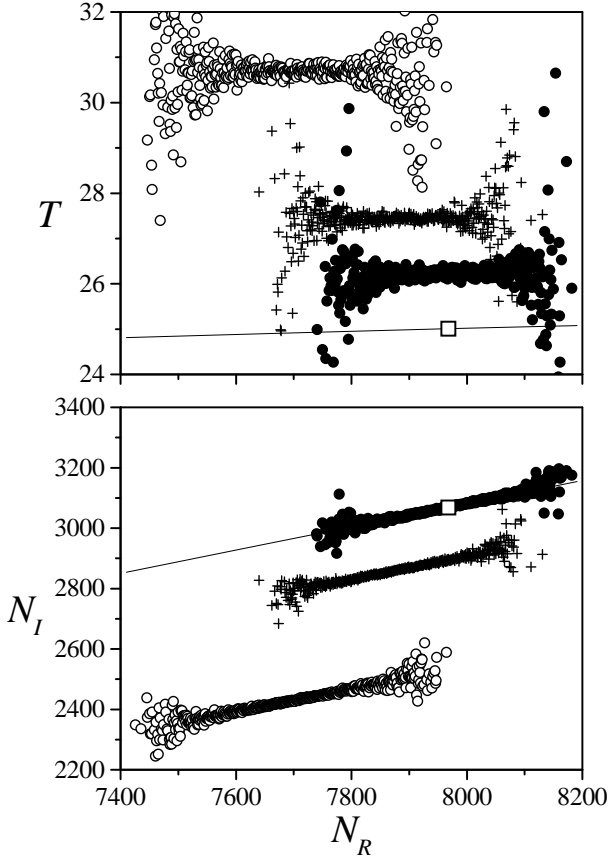}}
\end{center}
\caption{(a) The total evolution time $T$ and (b) the maximum number
of infected individuals $N_I$, as functions of the final
refractory population $N_R$, on a DSW with $N=10^4$ and $K=2$,
for $p=0.5$ (empty dots), $p=0.7$ (crosses), and $p=0.9$ (full
dots). Only the large-$N_R$ region, corresponding to the bump
structure, is shown. Data were extracted from $10^5$ realizations
for each value of $p$. In the cases where several  values of $T$
and $N_I$ were obtained for the same value of $N_R$, they were
averaged. Square symbols stand for the values of $N_I$ and $N_R$,
and the lower bound for $T$ predicted by the mean-field
approximation. The curves correspond to these same values for
varying $N$.}
\label{f9}
\end{figure}

The fact that propagation on DSWs is faster and more effective
than in SWNs becomes dramatically emphasized as soon as the
large-$N_R$ regime is analyzed. Figure \ref{f9} shows $T$ and
$N_I$ as functions of $N_R$ for large ``randomness,'' ranging
from $p=0.5$ to $0.9$.  These plots show the ``clouds''
corresponding  to the large-$N_R$ bump  structure, and are
therefore analogous  to the  inserts of  Figs. \ref{f3}b  and
\ref{f4}b. The typical  values of $N_R$,  $T$, and $N_I$  in
these DSW realizations  are  to  be  compared  with the
corresponding values for SWNs,  shown  in  Fig.  \ref{f6}.
While,  for  $N=10^4$, the fraction $r=N_R/N$ attains in SWNs a
maximum average level of about $0.44$  for $p=1$, in DSWs $r$
reaches a typical value close to $0.8$ for $p=0.9$. The
differences in $T$ and $N_I$ are even more drastic. In SWNs, their
extreme average values are $T\approx 75$ and $N_I\approx 440$,
whereas in DSWs they change to $T\approx 25 $ and $N_I \approx
3000$.

As the ``randomness'' of DSWs approaches its maximum value
$p=1$,  the system   should   be   satisfactorily   described
by   a  mean-field approximation.   The  analytical  treatment
of  our model within such approximation is developed in the
appendix. There we show that, for  a given size $N$, it is
possible to predict the values of $N_I$,  $N_R$, and a lower
bound for $T$.   These values are compared with  numerical
results  in  Fig.   \ref{f9}.   The  agreement  with  the
average  of numerical  results  is  very  good  for  the
largest  ``randomness,'' $p=0.9$.  These results, however, show
considerable fluctuations  with respect  to  the  values
predicted  by  the mean-field approximation. Moreover,
fluctuations  in   $N_R$,  $T$,  and   $N_I$  are   closely
correlated.  Note that the same kind of correlations were
suggested in SWNs by  the results  shown in  the inserts  of
Figs.   \ref{f3}b  and \ref{f4}b.  Since,  as discussed above,
DSWs can be  more efficiently implemented in numerical
experiments, our results  in Fig.   \ref{f9} correspond to a
number of realizations  considerably larger than  for SWNs, and
such correlations become apparent.

It turns out that the  correlations between the values of  $T$,
$N_I$, and $N_R$  obtained in  single realizations  can be
explained in terms of the mean-field approximation. In fact,
calculating $N_I$  and $N_R$, and the  lower bound for  $T$ from
this  approximation for {\it different}  values  of  the  system
size $N$---ranging from $N\approx 9200$  to  $10300$---we obtain
the  values  shown in Fig. \ref{f9}  as  curves. These values
successfully  reproduce   the correlation between  the three
quantities.  Specifically,  as   $N$ increases, $T$ decreases
slowly and  $N_R$ grows, while---as shown  in the
appendix---$N_I$  and  $N_R$  are  linearly correlated.  From a
phenomenological viewpoint, these results can be interpreted as
if  in each  individual   realization  the system  appears   to
have an ``effective'' size---close to, but different  from, its
actual size---plausibly determined by variations in the
effectiveness with which the rumor spreads over the population.

\section{Conclusion}

We have here studied the evolution of an epidemic-like model
evolving on small-world geometries.  The dynamics, which can be
interpreted  as the spreading of  a rumor, is  known to exhibit a
transition between regimes of localization and propagation at a
finite randomness of  the underlying disordered  geometry
\cite{z}.   Epidemiological models  on geometries that  plausibly
represent  social networks  and information webs---such as
small-world and scale-free networks \cite{bara}---have recently
attracted much attention, in  view of their potential role in  the
description  of  actual  risk  situations  associated  with
infectious diseases and computer viruses
\cite{ka,percol1,percol2,sato1,sato2}.

In our model, the effectiveness of propagation is characterized
by the total number $N_R$ of individuals  that have been infected
during  the whole evolution. Generally, in a single realization
of the process  on an asymptotically large system of size $N$,
propagation affects either a vanishingly small fraction of the
population, $N_R/N \approx 0$,  or a finite fraction $r$. For  a
randomness $p$ below the  critical point $p_c$ of the
localization-propagation transition, only the small-$N_R$ regime
is observed,  and the rumor  remains localized within  a limited
neighborhood  of  its  origin.  For  $p>p_c$,  on  the  other
hand, realizations  in   both  regimes   are  observed.   The
fraction   of realizations  in  the  large-$N_R$  regime,  in
fact,  grows  as  the randomness increases.

The dynamics of our model is completely described by the evolution
of the number $n_I$ of  infected individuals. A compact
characterization of this evolution  is given by  the total time
$T$ elapsed up  to the extinction of  the infected population,
the  maximum number $N_I$ of infected individuals at  a given
time,  and the total number $N_R$ of infected individuals during
the whole evolution. The effectiveness of propagation increases
when $T$ decreases, because spreading is faster, and when  $N_R$
and  $N_I$ grow, because the  rumor reaches a larger population.

Our results for small-world networks can be summarized as
follows. For any value of $p$, the small-$N_R$ regime is
characterized by power-law correlations between  $T$, $N_R$,  and
$N_I$.   It has been suggested  that these correlation could be
explained in terms of a  random-walk picture  of  the
propagation  process in the $n_I$-space. A rigorous analogy,
however,  can only  be achieved  in terms  of a biased random
walk with a rather complicated time-dependent bias. In the
large-$N_R$ regime,  the values  of  $T$,  $N_R$,  and  $N_I$
obtained  in single realizations are distributed around certain
typical values, which vary as the network randomness $p$ changes.
Specifically, as $p$ grows, $T$ decreases  and  both  $N_R$  and
$N_I$  increase, indicating that the propagation  process
becomes  increasingly effective.  The    three quantities show a
quite marked saturation as the randomness approaches its
limiting  value $p=1$.  The effectiveness  of propagation  is
also improved,  as expected,  when  the  average  number of
neighbors per individual grows.

We  have  also  studied  these  features  in a so-called dynamic
small world. Instead of considering  a frozen network of
interaction links, dynamics small worlds  admit that distant
contacts can occur  between any two individuals, chosen at random
at each evolution step. We  have here shown that, as  far as our
model  is concerned, propagation in  a dynamic small  world is
qualitatively the  same as  on a  small-world network. Namely,
the same kind of correlations between $T$, $N_R$, and $N_I$  and
the  same  dependence  with  $p$  observed  on small-world
networks are reproduced in dynamic small worlds. The main
quantitative difference between  both cases  is that  in
dynamics  small worlds the effectiveness of propagation  is
considerably higher.  Evolution times are  overall  shorter and
infected  populations   larger  than   on small-world networks.
In qualitative  terms, this is plausibly  due to the fact  that,
in  dynamics  small  worlds,  the  average number of interaction
partners per  individual is very  large and the effect of
backpropagation  is  comparatively  negligible, especially,  in
large populations.

To  our  knowledge,  this  is  the  first  time  that  the
evolution of a dynamical process is compared in detail on
small-world networks and in dynamic small worlds. It may be
conjectured that our main  conclusions regarding this comparison
will hold for a large class of processes.  A systematic
comparison  would  in  fact  be  desirable  since,  though
small-world networks have  attracted considerably more  attention
than dynamic  small  worlds,  the  latter  have  the  advantage
of  easier analytical  and  numerical  treatment  and, moreover,
provide a more realistic model of social systems with mobile
individuals.

\section*{Acknowledgements}

Fruitful discussion with G. Abramson, M. Kuperman, and L. G.
Morelli is gratefully acknowledged.

\appendix

\section*{Mean-field approximation}

Taking into  account that  the susceptible,  infected, and
refractory populations---which  we  denote  $n_S(t)$, $n_I(t)$,
and  $n_R(t)$, respectively---satisfy $n_S(t)+n_I(t)+n_R(t)=N$
for  all times, the mean-field evolution of our model is given by
the two equations
\begin{equation} \label{dnI}
\dot n_I = n_I\left(1-2\frac{n_I+n_R}{N} \right)
\end{equation}
and
\begin{equation} \label{dnR}
\dot n_R=n_I\frac{n_I+n_R}{N}.
\end{equation}
These equations can be implicitly solved in terms of the
auxiliary variable
\begin{equation}
s=\int_0^t n_I(t') dt'.
\end{equation}
We point out that the introduction of the variable $s$ in this
continuous approximation is  fully equivalent  to the  change
from  the real time scale to the measure of time  in evolution
steps used in the  discrete model.

With the initial conditions $n_I(0)=1$ and $n_R(0)=0$, the
solutions to Eqs. (\ref{dnI}) and (\ref{dnR}) are
\begin{equation} \label{nI}
n_I(s)=1-s+2(N-1)[1-\exp(-s/N)]
\end{equation}
and
\begin{equation} \label{nR}
n_R(s)=s-(N-1)[1-\exp(-s/N)].
\end{equation}
From these solutions, some relevant quantities can be immediately
calculated. The total number of steps $S$ needed for the
extinction of the infected population is the positive solution to
$n_I=0$, which corresponds to the transcendental equation
\begin{equation}
S-1=2(N-1)[1-\exp(-S/N)].
\end{equation}
For each value of $N$, this equation can be accurately solved by
numerical means. For asymptotically large $N$, it can be shown
that $S=k N$, where $k\approx 1.594$ is the positive solution to
$k=2[1-\exp(-k)]$ [note that $k=2r^*$; cf. Eq. (\ref{r*})]. The
final number of refractory individuals, $N_R$, can be evaluated
from Eq. (\ref{nR}) for $s=S$. This yields
\begin{equation}
N_R=\frac{S+1}{2}.
\end{equation}
Actually, this result holds not only in the mean-field
approximation, but for any value of $p$ and $N$ in both SWNs and
DSWs, as discussed in the main text.

The step $s_I$ at which the infected population $n_I(t)$ attains
its maximum $N_I$ is given by $\dot n_I=0$, i. e.  $s_I=N\ln
[2(N-1)/N]$. Replacing in Eq. (\ref{nI}) we obtain
\begin{equation}
N_I=N-1-N\ln \frac{2(N-1)}{N} \approx (1-\ln 2) N,
\end{equation}
where the right-hand side approximation holds for large $N$. In
this limit, combination of the above results makes it possible to
show that
\begin{equation}
N_I \approx \frac{2(1-\ln 2)}{k} N_R \approx 0.385 N_R.
\end{equation}

The  only  problematic  point  in  the  comparison  of  the
mean-field approximation with the original discrete model is the
evaluation of the total (real)  time $T$  elapsed up  to the
extinction of the infected population.  As  a  matter  of  fact,
in the mean-field approximation $n_I(t)$ decreases asymptotically
and vanishes only for $t\to  \infty$ (note that, however, this
limit corresponds to  a finite total  number of steps $S$).   The
total  time  $T$  must  therefore result from a plausible
definition using the mean-field results.  In the main text we use
the following criterion. We define $T$ as the time needed for
$n_I(t)$  to attain again its  original value $n_I=1$.  In other
words,  $T$ is the nontrivial solution  to $n_I(t)=1$,  which
can  be accurately obtained from numerical integration of  Eqs.
(\ref{dnI}) and (\ref{dnR}).  This gives a lower bound for the
actual total evolution time, that can  be directly compared with
our numerical results.

\end{multicols}

\end{document}